\documentclass[lettersize,journal]{IEEEtran}
\usepackage{amsmath,amsfonts}
\usepackage{algorithmic}
\usepackage{algorithm}
\usepackage{array}
\usepackage{textcomp}
\usepackage{stfloats}
\usepackage{url}
\usepackage{verbatim}
\usepackage{graphicx}
\usepackage{cite}
\usepackage{color}
\usepackage{latexsym}
\usepackage{makecell}
\usepackage{tipa}
\usepackage{amssymb}
\usepackage{booktabs}
\usepackage{amsmath}
\usepackage{float} 
\usepackage{subfigure}  
\usepackage{flushend}
\usepackage{booktabs}
\usepackage{tabularx}
\usepackage{multirow}
\usepackage{pifont}
\usepackage{makecell}
\usepackage{authblk}
\usepackage{svg}
\usepackage[justification=centering]{caption}
\begin{document}
\title{Efficient Semantic-aware Encryption for Secure Communications in Intelligent Connected Vehicles}

\author{Bizhu Wang, Zhiqiang Bian, Yue Chen, Xiaodong Xu, IEEE Senior Member,  \\  Chen Sun, IEEE Senior Member, Wenqi Zhang, Ping Zhang, IEEE Fellow 
\thanks{This work is supported by Beijing Natural Science Foundation No. L242012 and Sony (China) Research Laboratory  (Corresponding author: Xiaodong Xu, xuxiaodong@bupt.edu.cn).
Bizhu Wang, Zhiqiang Bian, Yue Chen, Xiaodong Xu and Ping Zhang are with the State Key Laboratory of Networking and Switching Technology, Beijing University of Posts and Telecommunications, Beijing, China; Xiaodong Xu and Ping Zhang are also with the Department of Broadband Communication, Peng Cheng Laboratory, Shenzhen, China; Chen Sun and Wenqi Zhang are with Sony (China) Research Laboratory.}}



\maketitle

\begin{abstract}

Semantic communication (SemCom) significantly improves inter-vehicle interactions in intelligent connected vehicles (ICVs) within limited wireless spectrum. However, the open nature of wireless communications introduces eavesdropping risks. To mitigate this, we propose the Efficient Semantic-aware Encryption (ESAE) mechanism, integrating cryptography into SemCom to secure semantic transmission without complex key management. ESAE leverages semantic reciprocity between source and reconstructed information from past communications to independently generate session keys at both ends, reducing key transmission costs and associated security risks. Additionally, ESAE introduces a semantic-aware key pre-processing method (SA-KP) using the YOLO-v10 model to extract consistent semantics from bit-level diverse yet semantically identical content, ensuring key consistency. Experimental results validate ESAE's effectiveness and feasibility under various wireless conditions, with key performance factors discussed.

\end{abstract}


\section{Introduction}

\begin{figure}[t]
\centering
\includegraphics[width=3.5in]{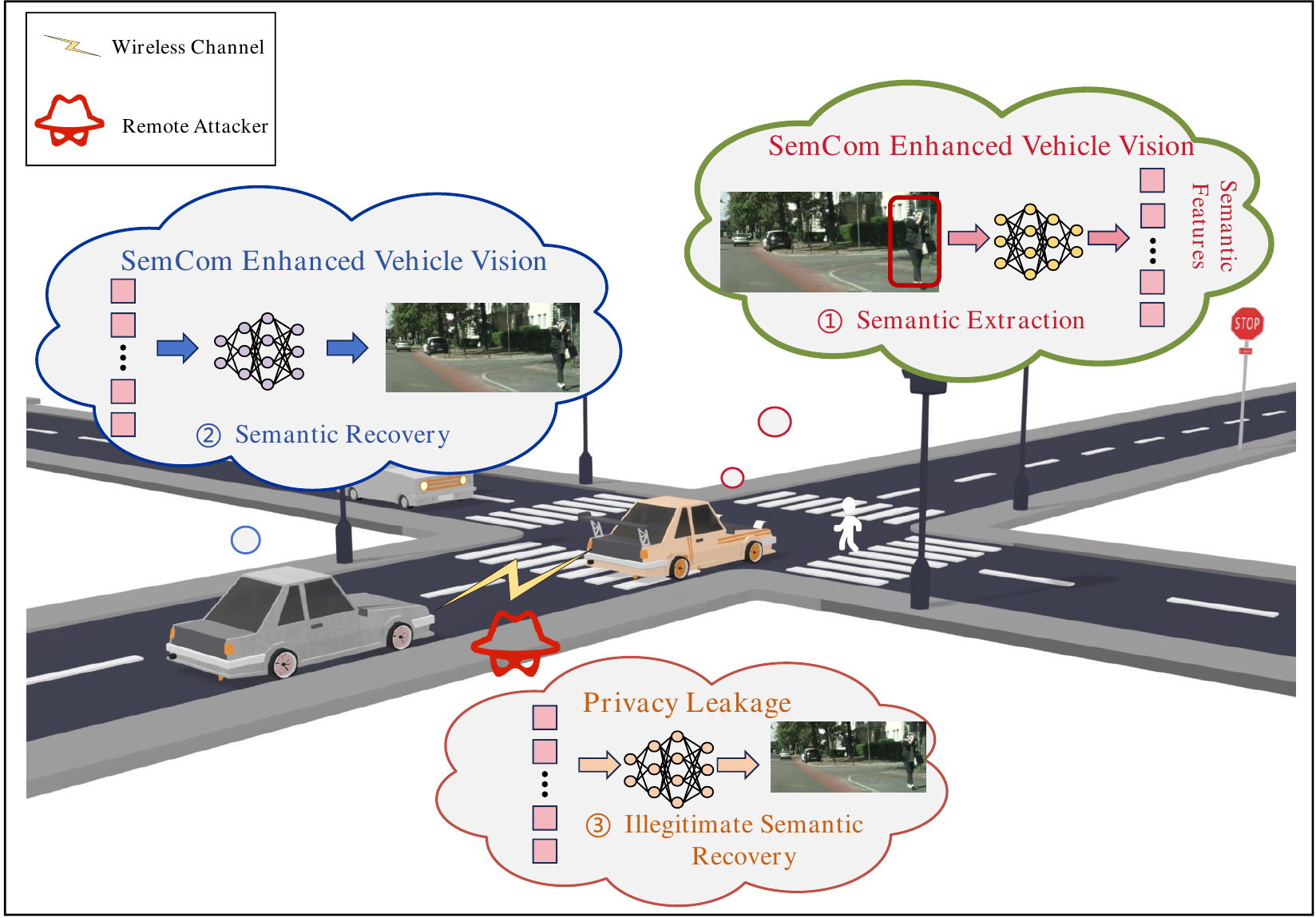}
\vspace{-0.4cm}
\captionsetup{font={footnotesize}}
\caption{Diagram of SemCom-enabled ICVs}
\label{framework}
\vspace{-0.6cm}
\end{figure}

\IEEEPARstart{T}{he} integration of semantic communication (SemCom) with intelligent connected vehicles (ICVs) promises to revolutionize vehicle interactions and boost road safety \cite{10598360}. By prioritizing the transmission of semantic information rather than raw images, SemCom-enabled ICVs not only share traffic data between vehicles but also reduce wireless spectrum usage, enhancing overall traffic services \cite{zhang2023semanticicv, xu2023semantic}. 
For example, as shown in Fig. \ref{framework}, the leading yellow vehicle monitors dynamic road conditions (e.g., pedestrians) and transmits semantic information to the following grey car, enabling it to detect blind spot traffic and adjust its driving accordingly.

Despite the significant benefits of SemCom-enabled ICVs, security remains a critical concern, particularly due to the vulnerability of wireless communications to eavesdropping \cite{qin2023securing}. As shown in Fig. \ref{framework}, the transmitted semantic information could expose sensitive data in ICVs, such as license plate and pedestrian facial details. Without proper safeguards, this exposed information can be exploited by malicious entities \cite{yang2024secure}.\par


Researchers have explored various solutions to secure semantic communication. For example, \cite{xu2024covert,10662946} uses neural network (NN) to generate artificial noise, maximizing the transmission rate while avoiding eavesdropper detection, but \cite{xu2024covert} assumes prior knowledge of the eavesdropper's channel conditions. Similarly, \cite{luo2023encrypted} employs NN-based encryption and decryption to allow valid nodes to reconstruct information and block eavesdroppers, though NN-based encryption and semantic similarity as security metrics lack wide recognition. Compared to NN-based methods, integrating cryptographic mechanisms into SemCom offers stronger security guarantees. For instance, \cite{tung2023deep} uses asymmetric key encryption, where the sender encrypts data with the receiver's public key and the receiver decrypts it with its private key, but this relies on pre-deployed keys and overlooks key distribution costs. To reduce key distribution complexity, \cite{qin2023securing,liu2023semprotector} combine physical layer encryption (PLE) with SemCom. Since PLE key generation rates drop to zero in static environments, \cite{qin2023securing,liu2023semprotector} generate semantic session keys using BLEU scores from historical interactions, encrypt the semantic key with the physical-layer key, and transmit them to the receiver. However, frequent session key transmissions increase communication overhead.

\begin{figure*}[t]
\centering
\includegraphics[width=7 in]{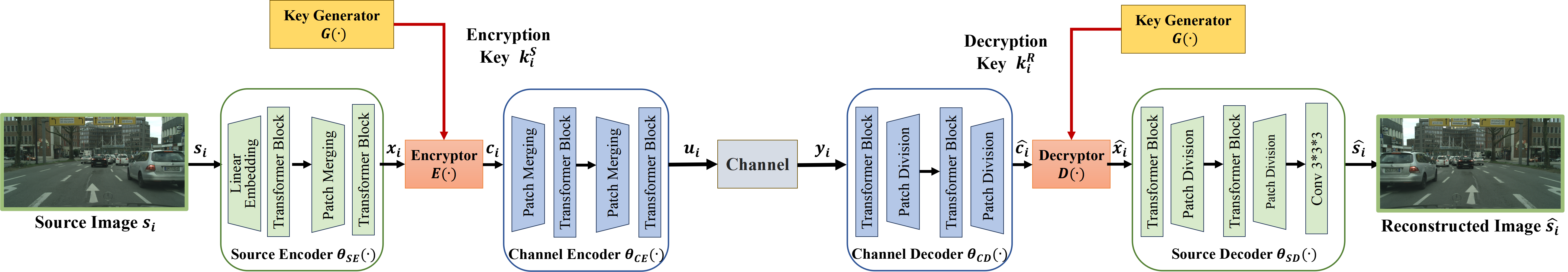}
\captionsetup{font={footnotesize}}
\caption{Flowchart of encryption-enabled SemCom ICVs}
\label{fig_2} 
\end{figure*}

Inspired by PLE, we identify the reciprocity between source and reconstructed information, which can be leveraged for efficient key generation. Following this, we propose the Efficient Semantic-aware Encryption (ESAE) mechanism to secure semantic communications in ICVs, addressing the limitations of existing methods. The key contributions are as follows:\par


\begin{itemize}
    \item We proposed the ESAE to secure the wireless semantic communications in ICVs, preventing eavesdropping while staying within wireless resource constraints.
    \item The proposed ESAE leverages the reciprocity of semantics in historical communications to generate consistent session keys at valid communication ends, thus eliminating the need to transmit encrypted session keys, reducing security risks and transmission costs.
    \item The proposed ESAE involves a novel Semantic-aware Key Pre-processing (SA-KP) method that utilizes the state-of-the-art object detection model, YOLOv10 \cite{wang2024yolov10realtimeendtoendobject}, to identify consistent semantics and enable key generation from semantically identical, yet bit-level diverse, previously communicated content at both ends.
    
    \item We introduced the Mean Consistency Rate of Semantic Key Generation (MCR-SKG) to evaluate key consistency and validated ESAE's security and feasibility through experiments, with key performance factors discussed.
\end{itemize}

\section{System Model}
\label{system}

Fig. \ref{fig_2} illustrates the framework of encryption-enabled SemCom ICVs. As assumed in \cite{luo2023encrypted}, the NN-based models in SemCom are public, ensuring consistent source-channel coding and decoding across all nodes. Mismatched models can be resolved using the synchronization algorithm from \cite{10553338}. In line with Kerckhoffs's principle, cryptosystem security relies on key secrecy rather than the encryption and key generation algorithms, so these algorithms are accessible to both ends. The semantic communication process is outlined below:


At the $i^{th}$ time-stamp, the semantic encoder $\theta_{SE}(\cdot)$ maps the source image $s_i \in S$ to a semantic feature stream $x_i = \theta_{SE}(s_i)$. The encryptor $E(\cdot)$ then encrypts the stream $x_i$ using the encryption key $k_{i}^{S} \in K$ generated by the key generator $G(\cdot)$, producing the encrypted semantic stream $c_{i} = E(x_{i}, k_{i}^{S})$. This stream is subsequently fed into the channel encoder $\theta_{CE}(\cdot)$, yielding the output $u_{i} = \theta_{CE}(c_{i})$.\par

Given the wireless channel coefficient $h$ and noise $n$, the received encrypted semantic stream at the receiver is expressed as $y_i = h \cdot u_i + n$. The channel decoder $\theta_{CD}(\cdot)$ processes $y_i$, yielding $\hat{c_i} = \theta_{CD}(y_i)$. The receiver then inputs $\hat{c_i}$ and the decryption key $k_{i}^{R} \in K$ generated by $G(\cdot)$ into the decryptor $D(\cdot)$, resulting in $\hat{x_i} = D(\hat{c_i}, k_{i}^{R})$. Finally, the semantic decoder $\theta_{SD}(\cdot)$ reconstructs the image as $\hat{s_i} = \theta_{SD}(\hat{x_i})$.\par

Based on the above system model, the objectives of the SemCom-enabled ICVs are twofold:\par

1) The semantic error, defined as the difference between ${s_i}$ and $\hat{s_i}$, should be minimized to approach an arbitrarily small value $\varepsilon_{thre}$. The objective function can be modeled as
\begin{equation}
    \text{T1:}~\mathcal{L} = \sum_{s}\mathcal{F}({s},{\hat{s}}) < \varepsilon_{thre}
    \label{eq:T1}
\end{equation}

where similarity measure $\mathcal{F}(\cdot)$ can be assessed using various metrics, such as the Peak Signal-to-Noise Ratio (PSNR).\par
2) An eavesdropper Eve cannot infer any details about the plain-text image $s_i$ from the encrypted semantic stream even when using standard semantic and channel encoders and decoders. The objective can be modeled as 
\begin{equation}
    \text{T2:}~adv_{E,D} = 2\mathcal P(b'=b)-1< \varepsilon_{thre}
    \label{eq:T2}
\end{equation}

where $E$ and $D$ represent the encryption and decryption functions. $b$ is sampled from a Bernoulli distribution with a success probability of 0.5. Eve chooses two plain-text $s_0$ and $s_1$ of equal dimension from $S$ and provides them to an oracle, who computes cipher-text $c=E(s_b,k)$ with the key $k$. Then, the oracle returns the cipher text $c$, which is one of the chosen messages determined by $b$. Based on $c$, Eve tries to determine $s_b^{'}$ as the plain text used to compute $c$. Thus, an advantage approaching $\varepsilon_{thre}$, in other words, $P(b'=b)=0.5$, indicates that Eve's guess is no better than random.

\begin{figure*}[t]
\centering
\includegraphics[width=6.0 in]{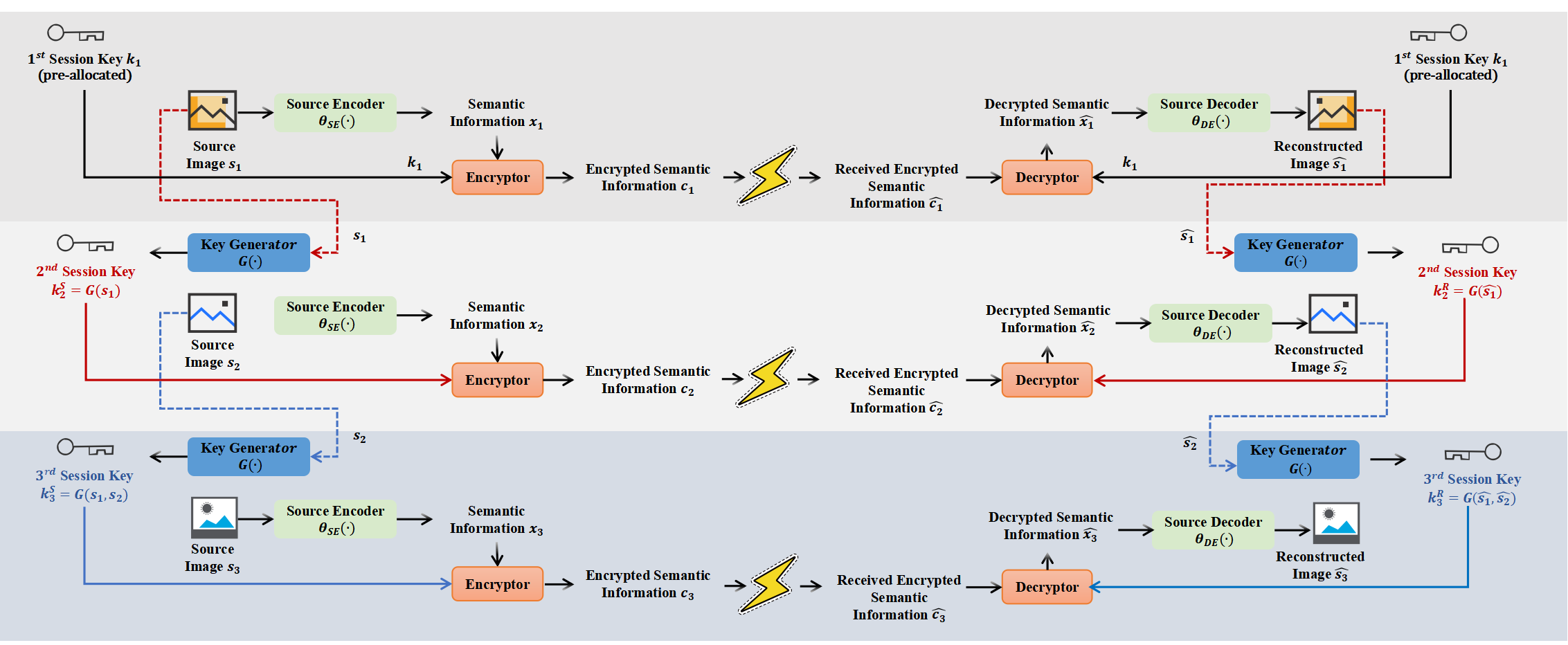}
\captionsetup{font={footnotesize}}
\caption{Procedure for semantic-aware session key generation and update in ESAE (channel encoding/decoding omitted for simplicity)}
\label{fig_3}
\end{figure*}

\section{Proposed ESAE Mechanism}
\label{method}
To achieve the objective in (\ref{eq:T1}), the semantic-related modules ${\theta_{SE}}, {\theta_{CE}}, {\theta_{CD}},$ and ${\theta_{DE}}$ are trained to minimize the semantic error $\mathcal{L}$. The objective in (\ref{eq:T2}), can be met using traditional cryptographic algorithms, but these introduce complex key management and communication overhead due to frequent key distribution. To address this, we propose ESAE, which enables communication parties to derive identical session keys from previously communicated semantics, eliminating the need for complex key management and frequent key distribution. The principles of ESAE are illustrated in Fig. \ref{fig_3}.
\begin{figure*}[h]
\centering
\includegraphics[width=6.0 in]{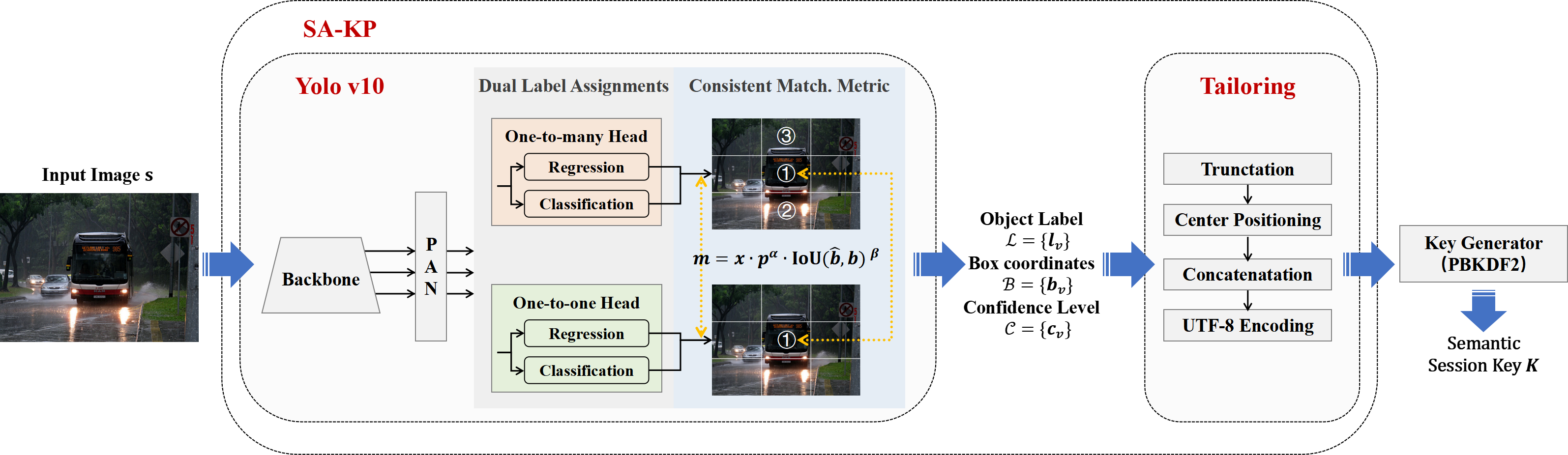}
\captionsetup{font={footnotesize}}
\caption{Diagram of SA-KP for semantic key generation}
\label{preprocess}
\end{figure*}

\paragraph{\textbf{Initialization Phase}} ESAE applies asymmetric cryptography to securely exchange the initial session key between communicating ends at the very beginning. We assume this process is secure, with the first session key denoted as $k_1$.
\paragraph{\textbf{Initial Secure Semantic Transmission}} Based on the system model in Section~\ref{system}, the source node encrypts the semantic information $x_1$ of the first image $s_1$ with the first session key $k_1$, resulting in $c_{1}=E(x_{1},k_{1})$. The receiver, having access to $k_1$, decrypts it to obtain the semantic information $\hat{x_1}=D(\hat{c_{1}},k_{1})$, reconstructing the image as $\hat{s_1}$.

\paragraph{\textbf{Semantic Session Key Generation and Update}} 
Then, ESAE generates identical session keys at communication ends using previously communicated semantics, preventing eavesdropping. In detail, the sender derives the second session key $k_{2}^{S}$ from the previously communicated content $s_1$, as $k_{2}^{S} = G(s_1)$, while the receiver derives $k_{2}^{R}$ from the reconstructed image $\hat{s_1}$, as $k_{2}^{R} = G(\hat{s_1})$. Password-Based Key Derivation Function 2 (PBKDF2) can be used in the key generator $G(\cdot)$.\par

From the above observations, 
the core of ESAE is ensuring both ends generate identical keys from $s_1$ and $\hat{s_1}$, so that $k_{2}^{S} = k_{2}^{R}$. \textbf{Since $s_1$ and $\hat{s_1}$ are semantically, but not bit-level, identical, generating consistent keys from semantically similar images is crucial, as addressed in Section~\ref{semkey}}.

\paragraph{\textbf{Subsequent Secure Semantic Transmission}}
When transmitting the second image $s_2$, the sender encrypts the corresponding semantic information $x_2$ using the latest session key $k_{2}^{S}$, resulting in $c_2=E(x_2,k_{2}^{S})$. Then, the decryption process at the receiver can be modeled as  $\hat{x_2}=D(\hat{c_2},k_{2}^{R})$. When $k_{2}^{S}=k_{2}^{R}$, then the recovered image after semantic decoding is $\hat{s_2}$. Here, AES (Advanced Encryption Standard) can be applied by the encryptor and the decryptor denoted as $(E,D)$ described in Section~\ref{system}.\par
Similar to the second semantic transmission, the j\textsuperscript{th} session key generated by the sender and the receiver can be denoted as $k_{j}^{S}=G(s_1,s_2,\cdots,s_{j-1})$ and $k_{j}^{R}=G(\hat{s_1},\hat{s_2},\cdots,\hat{s_{j-1}})$. When facing computational or buffer constraints, a window size $\mathcal{T}$ can be applied during key generation. Specifically, the session key $k_j$ is derived from the most recent $\mathcal{T}$ communicated contents, denoted as $k_{j}^{S}=G(s_{j-\mathcal{T}},s_{j-\mathcal{T}+1},\cdots,s_{j-1})$. The choice of $\mathcal{T}$ should balance security and efficiency.\par

\section{Proposed SA-KP Mechanism}
\label{semkey}

As mentioned above, the core of ESAE is to enable both sender and receiver to derive the same session key from semantically similar, yet bit-level different, previously communicated images. To achieve this, we propose the SA-KP mechanism, which integrates the new benchmark object detection model, YOLOv10 \cite{wang2024yolov10realtimeendtoendobject}, allowing both ends to identify identical semantics from bit-level diverse communicated content.

As shown in Fig. \ref{framework}, the input image $s$ first passes through a backbone network for multi-level feature extraction, followed by multi-scale feature fusion through the Path Aggregation Network (PAN), which enhances object detection across various sizes. YOLOv10 employs a dual-label assignment mechanism with two detection heads: One-to-many for broad predictions and One-to-one for precise predictions, handling both bounding box regression and classification. The predictions are evaluated using a consistent matching metric $m=x\cdot p^{\alpha}\cdot IoU(\hat{b},b)$. Here, $p$ is the classification score, $\hat{b}$ and $b$ denote the predicted and ground-truth bounding boxes, respectively. $x$ indicates the spatial prior, determining whether the anchor point of the prediction is within the instance. $\alpha$ and $\beta$ are two hyper-parameters that balance the impact of semantic prediction task and the location regression task. Finally, YOLOv10 outputs the object labels $\mathcal{L}$, bounding box coordinates $\mathcal{B}$, and confidence scores $\mathcal{C}$. 

Then, the YOLOv10 outputs are tailored to specific computational, storage, or security needs by selecting only the top $K$-confidence objects. To ensure consistent key generation at both communication ends, the image is divided into $Q$ grids, with each top K-confidence object's bounding box $\mathcal{B}$ center assigned to its respective grid. The selected object labels $\mathcal{L}$ and grid indices are first sorted based on the grid indices. After that, they are encoded as a UTF-8 byte sequence, and fed into the key generator to produce the encryption key $k$.\par

In conclusion, by properly setting the window sizes $\mathcal{T}$, the number of top-confidence objects $K$ and the grid division $Q$, SA-KP ensures identical keys can be independently generated from the source and reconstructed images, which are semantically consistent but differ at the bit level.

\section{Performance Evaluation}

From its design, ESAE clearly demonstrates the ability to reduce key transmission overhead and management complexity. This section further assesses its feasibility and security. In detail, Section~\ref{metric} introduces a key consistency metric, which forms the basis for the feasibility simulations and performance analysis in Section~\ref{results}. Section~\ref{security_analysis} evaluates security by analyzing key search spaces at both the bit and semantic levels.
\subsection{Simulation Settings}
\subsubsection{Data set} 

We utilized the Common Objects in Context (COCO) dataset \cite{lin2014microsoft}, a large-scale and widely-used dataset in computer vision. With 330,000 images and 2.5 million labeled instances across 80 object categories, COCO is ideal for tasks such as object detection and segmentation, emphasizing scene complexity and object-context relationships. COCO includes extensive ICVs-related imagery, featuring cars, buses, trucks, and traffic lights, primarily captured in outdoor environments.

\subsubsection{ESAE implementation}
Refer to Fig. \ref{fig_2}. ESAE operates on the advanced NTSCC framework \cite{9791398}, which is compatible with the COCO dataset across various wireless communication scenarios. The Yolov10 \cite{wang2024yolov10realtimeendtoendobject} used by SA-KP is fine-tuned to ensure alignment between the outputs when processing the source and the reconstructed information from NTSCC. Besides, key generator applies PBKDF2 and while AES-128 is used for encryption. 

\subsection{Evaluation Metrics}
\label{metric}
It is crucial that the semantic keys generated at both communication ends achieve an exact bit-level match, expressed as $k^{S} = k^{R}$. Let $\mathbb{E}[\cdot]$ denote the expectation operator and $\mathbb{I}$ the indicator function, which equals 1 if $k^{S} = k^{R}$ and 0 otherwise. The Mean Consistency Rate of Semantic Key Generation (MCR-SKG) is defined as:

\begin{equation}
\text{MCR-SKG} = \mathbb{E}[\mathbb{I}(k^{S} = k^{R})]
\end{equation}

\subsection{Security Analysis}
\label{security_analysis}
For each encrypted semantic communication, the bit-level key search space is $2^{128}$. At the semantic level, the key search space is calculated as $\frac{{(NQ)}^{K\mathcal{T}}}{K!}$, where $N,Q,K,\mathcal{T}$ denote the total number of object classes, the number of grid division, chosen number of top-confidence objects and the predefined window size, respectively. Thus, the semantic key search space surpasses $2^{128}$ when 
$KTln(NQ)>128ln2+ln(K!)$, ensuring that ESAE is semantically secure.\par

Moreover, ESAE enhances security by generating keys independently at each communication end, reducing risks associated with key transmission. In contrast, the semantic encryption scheme mentioned in \cite{qin2023securing} is prone to cascading vulnerabilities in static environments. An attacker who breaches one encrypted transmission, gaining access to both the key and plain-text, can calculate the BLEU score and anticipate subsequent keys, thus decrypting future communications. However, in ESAE, an attacker must compromise $\mathcal{T}$ encrypted semantic communications timely and simultaneously to succeed.\par

\subsection{Simulation Results and Analysis}
\label{results}

\begin{figure}
    \centering
     \subfigure[SA-KP outputs on the sender]{\includegraphics[width=0.47\linewidth]{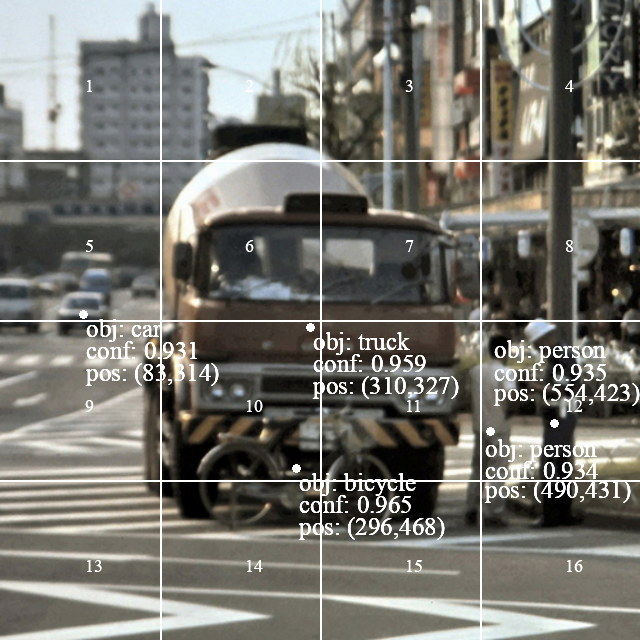}}\hspace{1mm}
    \subfigure[SA-KP outputs on the receiver]{\includegraphics[width=0.47\linewidth]{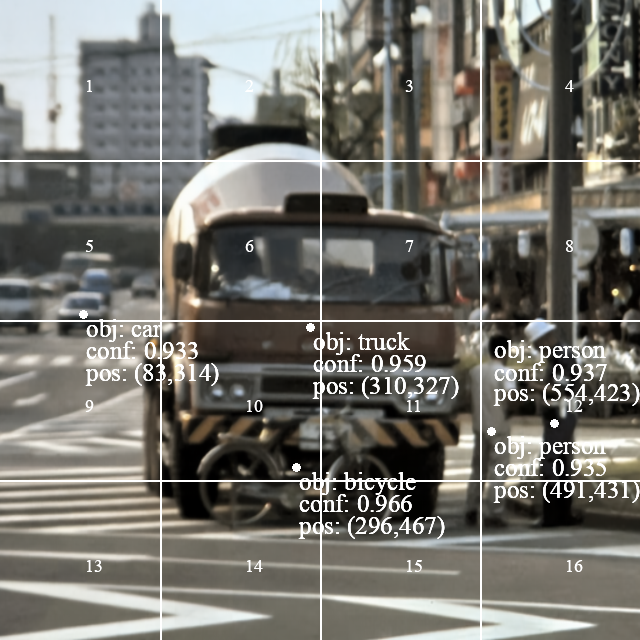}}\hspace{1mm}
   
    \vspace{-0.2cm}
    \captionsetup{font={footnotesize}}
    \caption{The visualization of SA-KP outputs when $\mathcal{T}=1$ and SNR=25 dB}
    \label{visual}
\end{figure}

\begin{figure*}
    \centering

     \subfigure[SNR=5dB]{\includegraphics[width=0.30\linewidth]{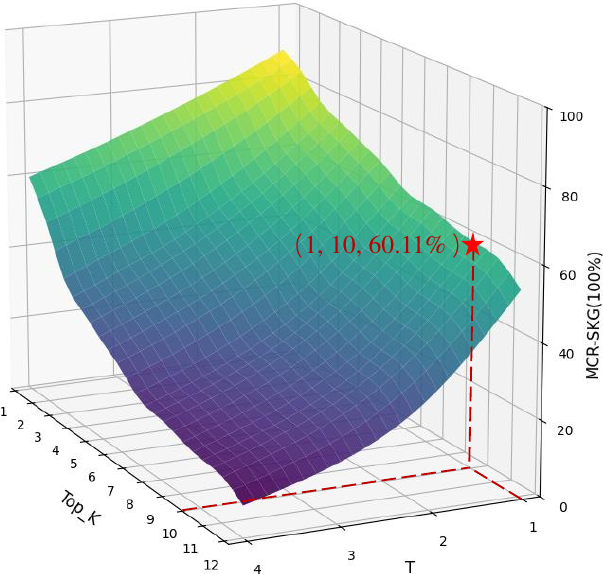}}\hspace{1mm}
    \subfigure[SNR=10dB]{\includegraphics[width=0.30\linewidth]{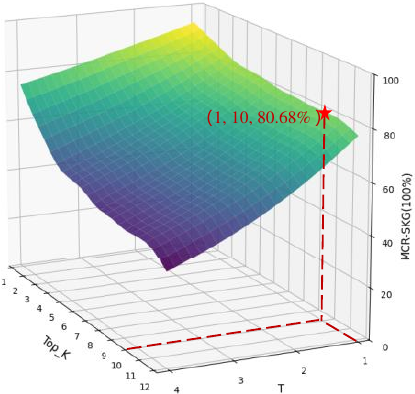}}\hspace{1mm}
    \subfigure[SNR=25dB]{\includegraphics[width=0.36\linewidth]{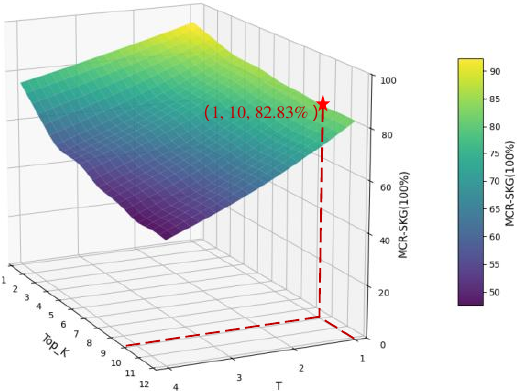}}\hspace{1mm}
    \vspace{-8pt}
    \caption{MCR-SKG comparisons across various SNR, $K$ and $\mathcal{T}$ when $Q=900$}
    \label{valueresults}
\end{figure*}

Figs. \ref{visual}a and b present the visual outputs of the SA-KP when the source inputs the t-th transmitted image and the receiver inputs the corresponding t-th received image at $\mathcal{T}=1$ and SNR=25 dB. Despite the presence of bit-level differences, which are observable from the variations in confidence values and grid locations of the bicycle object, the SA-KP modules of both the sender and the receiver yield the same top-ranked object categories and object center grid indices. In line with the design of ESAE, by feeding these extracted semantics (the top 5 pairs of object and grid index) into PBKDF2 to update the session keys for the next round, the same encryption and decryption keys will be generated. This effectively demonstrates the efficacy of ESAE in ensuring key consistency and facilitating secure semantic communication.


\begin{figure}[t]
\centering
\includegraphics[width=0.82\linewidth]{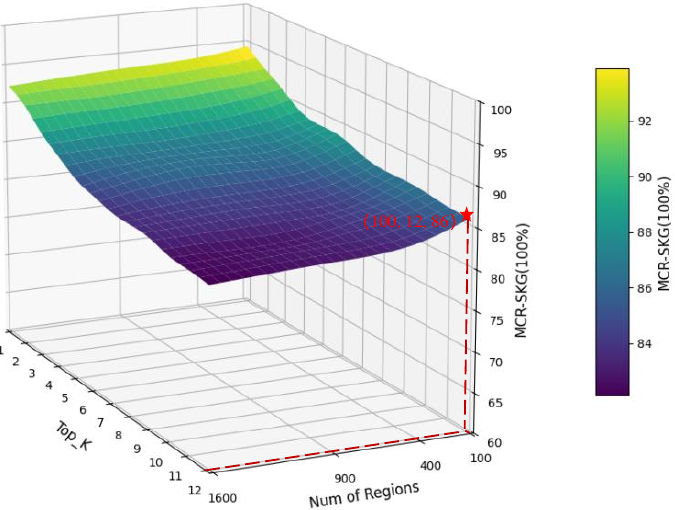}
\captionsetup{font={footnotesize}}
\caption{MCR-SKG comparisons across various $K$ and $Q$ when $\mathcal{T}=1$}
\label{valueresults2}
\vspace{-0.3cm}
\end{figure}

Fig. \ref{valueresults} states that 1) MCR-SKG exhibits an upward trend as SNR increases from 5 dB to 25 dB since enhanced information recovery heightens the probability of obtaining consistent results from the SA-KP process applied to the source image and the re-constructed image. The pentagrams on the figures represent the highest MCR-SKG when the semantic-level search space surpasses the bit-level search space, specifically when it is greater than $2^{128}$, as discussed in Section~\ref{security_analysis}. 2) Besides, larger values of $K$ and $\mathcal{T}$ tend to reduce MCR-SKG. A large $K$ can lead to inaccuracies in the extraction of low-confidence objects. Meanwhile, a large $\mathcal{T}$ may initiate a chain reaction of decryption errors, thereby resulting in inconsistent keys in subsequent rounds. Thus, choosing appropriate $K$ and $\mathcal{T}$ values is essential for balancing security and feasibility.\par

Besides, the choice of $Q$ also matters. As shown in Fig. \ref{valueresults2}, when the number of regions (grids) are varied, the highest MCR-SKG can further increase to 86$\%$ at SNR=25 dB and $\mathcal{T}=1$, while meeting the requirements of semantic-level security. 

In conclusion, the upper limit of MCR-SKG hinges on whether SA-KP can extract identical semantic information from transmitted historical images and received recovered ones, which is closely tied to hyper-parameter selection, object detection model performance, and the wireless environment. Thus, enhancing the object detection model's performance to effectively handle the challenges posed by SNR and lossy transmissions to semantic information extraction is highly beneficial for boosting the feasibility of ESAE.

\section{Conclusion}
This paper presents the ESAE to efficiently secure semantic communications in ICVs by utilizing the reciprocity of historical transmitted and received contents to independently generate session keys at the communication ends, thereby reducing the overhead associated with key transmission. Then, the SA-KP method is proposed to ensure consistent keys from semantically identical but bit-wise different content. Simulations validate ESAE's security and feasibility, emphasizing the need for well-adjusting the optimal value of window size, the number of top-confidence objects and the grid divisions to balance security and key generation consistency.

\bibliographystyle{IEEEtran}
\bibliography{IEEEabrv,references}


\vfill

\end{document}